\documentclass[pra,showpacs,twocolumn]{revtex4}
\usepackage{ulem}
\usepackage{amsmath}
\usepackage{amstext}
\usepackage{latexsym}
\usepackage{graphicx}
\usepackage{amsfonts}
\usepackage{epsfig}

\begin{document}

\title{Coherent-state phase concentration by quantum probabilistic amplification}
\author{Petr Marek and Radim Filip}
%\footnote{email:filip@optnw.upol.cz, tel:+420-68-5631572,
%fax:+420-68-5224246}
\affiliation{Department of Optics, Palack\' y University,\\
17. listopadu 50,  772~07 Olomouc, \\ Czech Republic}
\date{\today}
\begin{abstract}
We propose novel coherent-state phase concentration by probabilistic measurement-induced amplification. The amplification scheme uses novel architecture, thermal noise addition (instead of single photon addition) followed by feasible multiple photon subtraction using realistic photon-number resolving detector. It allows to substantially amplify weak coherent states and simultaneously reduce their phase uncertainty, contrary to the deterministic amplifier.
\end{abstract}
\pacs{03.67.Hk, 03.67.Dd}

\maketitle

Quantum optics has an extraordinary capability to combine observations of both wave and particle phenomena. For example, a coherent signal from a laser can be measured both by phase-sensitive interferometric detectors and by the particle sensitive photon-counting \cite{Glauber}. The coherence of light allows for encoding of information into phase of the field, instead just into intensity. However, this approach requires operations with high level of coherence, and therefore classical processing methods based on measurement and re-preparation are not suitable \cite{andfil}. The difference is even more pronounced, if the optical signals carrying phase information have very weak intensities, as in this case is the applicability of classical methods even further reduced by effects of loss and quantum noise. The main aim of the processing of phase information carried by quantum states is to reduce the noise and amplify the phase information to compensate for loss. In general, we seek to enhance an unknown phase of an optical signal by deterministic or probabilistic methods, where the aim of probabilistic procedures is to qualitatively overcome the limits given by deterministic operations.

Weak coherent states are non-orthogonal, very strongly overlapping if the amplitude is small, which can happen after strong attenuation. Therefore, it is highly desirable to re-amplify the states, but in a way that improves the phase information. It can be shown, the Gaussian phase-insensitive amplification \cite{Gamp,ulr}, makes phase information only worse due to fundamental quantum noise penalty \cite{ampnoise}.

On the other hand, a possible probabilistic noiseless amplifier making transformation $|\alpha\rangle\rightarrow |g\alpha\rangle$, where the gain $g>1$, can increase amplitude of the coherent state and subsequently concentrate the phase information. One way to implement this operation relies on the quantum scissors approach, limiting the dimension of the used Hilbert state \cite{ampi1}. The input coherent state is split into $M$ weak copies, which can be approximated by $(|0\rangle+\alpha/M|1\rangle+\cdots)^{\otimes M}$ and probabilistically amplified to $(|0\rangle+g\alpha/M|1\rangle)^{\otimes M}$. For a small value of $\alpha/M$ the following Gaussifying concentration yields a finite Hilbert space approximation of $|g\alpha\rangle$. However, the procedure requires multiple single photon sources perfectly coherent with the input state and high interferometric stability of the multi-path interferometer. Another approach is based on still highly sophisticated cross-Kerr nonlinearity at a single photon level followed by homodyne detection \cite{ampi2}. This kind of amplifier scheme was already suggested in Ref.~\cite{Fiu2} to concentrate entanglement.

In this Letter, we propose a conceptually novel scheme for concentration of phase of coherent states using a probabilistic highly nonlinear amplifier. Our method is based on adding thermal noise to the unknown coherent state followed by multiple photon subtraction using photon-number resolving detector. This procedure probabilistically amplifies coherent state, increasing its mean photon number and simultaneously substantially reducing the phase noise. It results in probabilistic concentration of phase information, which is impossible using Gaussian operations. Remarkably, the scheme requires neither single photon sources nor high interferometric stability. As the resource for the highly non-linear amplification serves the continual noise modulation of the signal mode. %The multiple photon subtraction can be implemented using linear coupling to the signal followed by a photon number resolving detector \cite{detect}.

The quality of information carried by the phase is difficult to asses, as phase is not a quantum mechanical observable and therefore it cannot be directly and ideally measured. However, each measurement devised to divine the phase of the state can be characterized by a real positive-semidefinite matrix $H$, which is used in obtaining the phase distribution $P(\theta) = \mbox{Tr}[\rho F(\theta)]$, where $F(\theta) = 1/2\pi \sum_{m,n=0}^{\infty}\exp(i\theta(m-n))H_{mn}|m\rangle\langle n|$ \cite{Wis}. The actual form of the matrix $H$ depends on the process used to extract the phase information. For example, for phase obtained by the the most common heterodyne measurement, consisting of a balanced beam splitter and a pair of homodyne detectors measuring conjugate quadratures, the matrix elements are $H_{mn} = \Gamma[(n+m)/2 +1]/\sqrt{n!m!}$. Ultimately, for the ideal canonical phase measurement $H_{mn} =1$, for which the $F(\theta)$ is a projector on the idealized phase state $|\theta\rangle = \sum_{n=0}^{\infty}e^{i\theta n}|n\rangle$. To obtain a single parameter characterizing the phase, we can use the distribution to calculate the phase variance $V = |\mu|^{-2} -1$, where $\mu=\langle \exp(i\theta)\rangle$ and subscripts $H$ and $C$ will be used to distinguish between the heterodyne and the canonical measurements, respectively. For calculations of the canonical measurement we can simply use formula $\langle \exp(i\theta)\rangle=\int_{-\pi}^{\pi}P(\theta)\exp(i\theta)d\theta=\mbox{Tr}\sum_{n=0}^{\infty}|n\rangle\langle n+1|\rho$. %There are also other definition of the phase variance in the terms of $\mu$ \cite{othervar}.

The coherent states can be expressed as $|\alpha\rangle = \exp(-|\alpha|^2/2)\sum_{n=0}^{\infty} \alpha^n/\sqrt{n!} |n\rangle$. For these states, the quality of phase encoding is fully given by the mean number of coherent photons $N=|\alpha|^2$ and the phase variance obtained using \cite{Wis}
\begin{eqnarray}\label{hol1}
    \mu_{C} &=& e^{-|\alpha|^2}\alpha\sum_{n=0}^{\infty} \frac{|\alpha|^{2n}}{n!\sqrt{n+1}},\nonumber\\
    \mu_{H} &=& e^{-|\alpha|^2}\alpha\,_1F_1
    \left(\frac{3}{2};2;|\alpha|^2\right)\frac{\Gamma\left[\frac{3}{2}\right]}{\Gamma\left[2\right]}
\end{eqnarray}
are both monotonously decreasing function of the mean photon number $N$. For weak coherent states with $N<1$, the variances can be well approximated by
\begin{eqnarray}\label{variances}
 V_{C}(N) &\approx & N^{-1}+1-\sqrt{2}+\mbox{O}^2[N], \nonumber \\
 V_{H}(N) &\approx& 4/(\pi N) + (-1 + 2/\pi)+\mbox{O}^2[N],
\end{eqnarray}
if we take only dominating terms into account. In the following, we will focus primarily on the canonical phase variance, going back to the heterodyne detection only later.

The deterministic phase-insensitive (Gaussian) amplifier \cite{Gamp} increases the phase variance of coherent state.
To prove this, we can use similar method as in \cite{Aspachs} to calculate
\begin{equation}\label{mu_ampg}
\mu_{C}=\frac{\alpha^*}{\pi}\int_{0}^{\frac{1}{G}}\frac{\exp\left(-xGN\right)}
{\sqrt{-\ln{\frac{1-Gx}{1-(G-1)x}}}} dx ,
\end{equation}
where $G = g^2$ is the linear amplification gain. We can now use (\ref{mu_ampg}) to obtain the phase variance and the original statement can be verified numerically. 

On the other hand, the probabilistic noiseless amplification $|\alpha\rangle\rightarrow |g\alpha\rangle$ improves the amplitude while preserving the coherent nature of the state, thus reducing the phase variance.
However, there is another mechanism that can be employed to this end. Consider the single photon addition (described by $a^{\dagger}|n\rangle=\sqrt{n+1}|n+1\rangle$) followed by the single photon subtraction (described by $a|n\rangle=\sqrt{n}|n-1\rangle$) applied to a weak coherent state (approximately, $|\alpha\rangle=|0\rangle+\alpha|1\rangle$). This corresponds to $aa^{\dagger}(|0\rangle+\alpha|1\rangle)\rightarrow a(|1\rangle+\sqrt{2}\alpha|2\rangle)\rightarrow |0\rangle+2\alpha|1\rangle$. For low $N$ this reduces the phase variance approximately by a factor of four. Note, the canonical variance actually decreases in both the creation and the annihilation process.

%The reduction arises from the fundamental quantum mechanical noise, just imagine a classical analog of the creation doing just $a^{\dagger}|n\rangle=|n+1\rangle$ and the reduction of the phase variance will disappear.

%%%%%%%%%%%%%%%%%%%%%%%%%%%%%%%%%%%%%%%%%%%%%%%%%%%%%%%%%%%%%%%%%%%%%%%%%%%%%%%%%
\begin{figure}
\centerline{\psfig{figure = 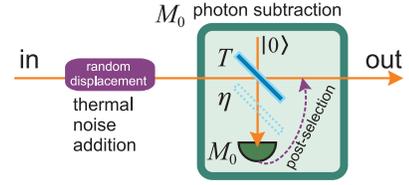,width=0.6\linewidth}}
\caption{Scheme for the phase concentration by probabilistic amplification of a coherent state.}\label{fig_setup}
\end{figure}
%%%%%%%%%%%%%%%%%%%%%%%%%%%%%%%%%%%%%%%%%%%%%%%%%%%%%%%%%%%%%%%%%%%%%%%%%%%%%%%%%%%%
%%%%%%%%%%%%%%%%%%%%%%%%%%%%%%%%%%%%%%%%%%%%%%%%%%%%%%%%%%%%%%%%%%%%%%%%%%%%%
\begin{figure}
\textbf{(a)}\hskip3.5cm \textbf{(b)}\vskip-0.5cm
\centerline{\psfig{figure = 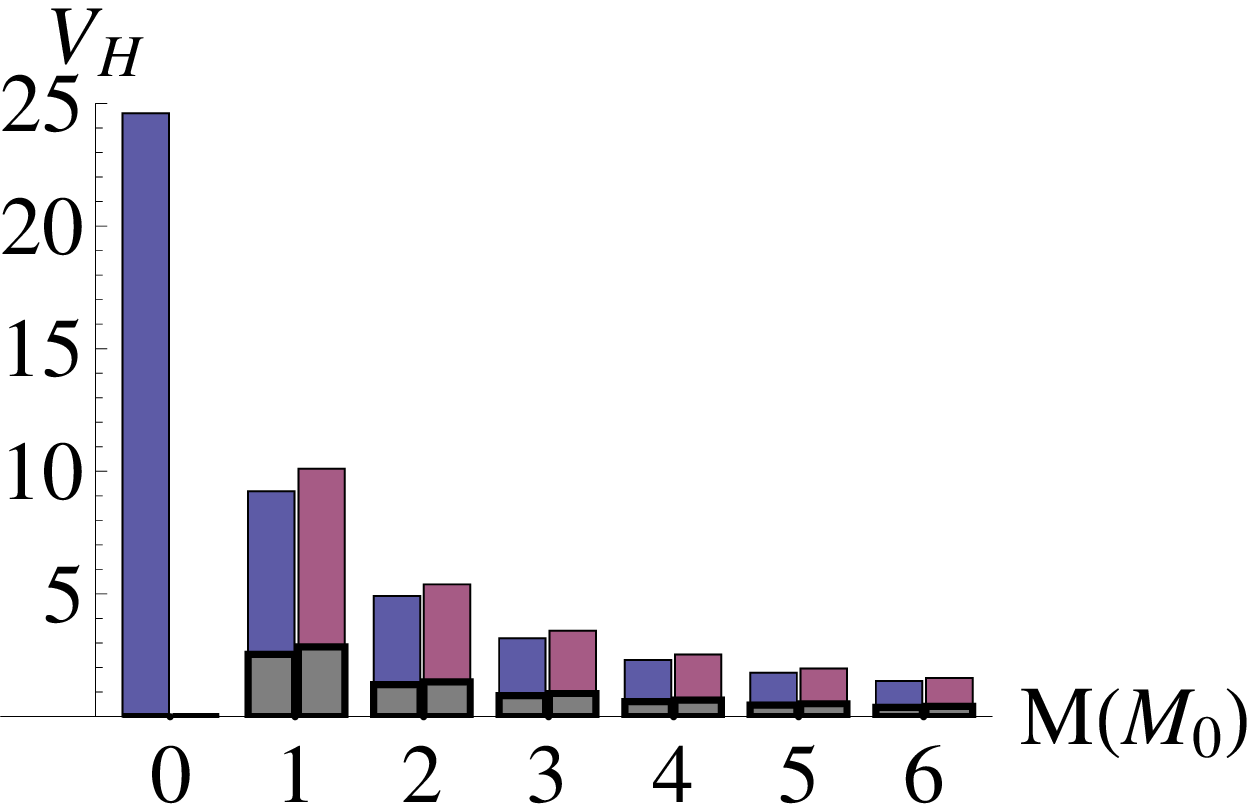,width=0.45\linewidth}\psfig{figure = 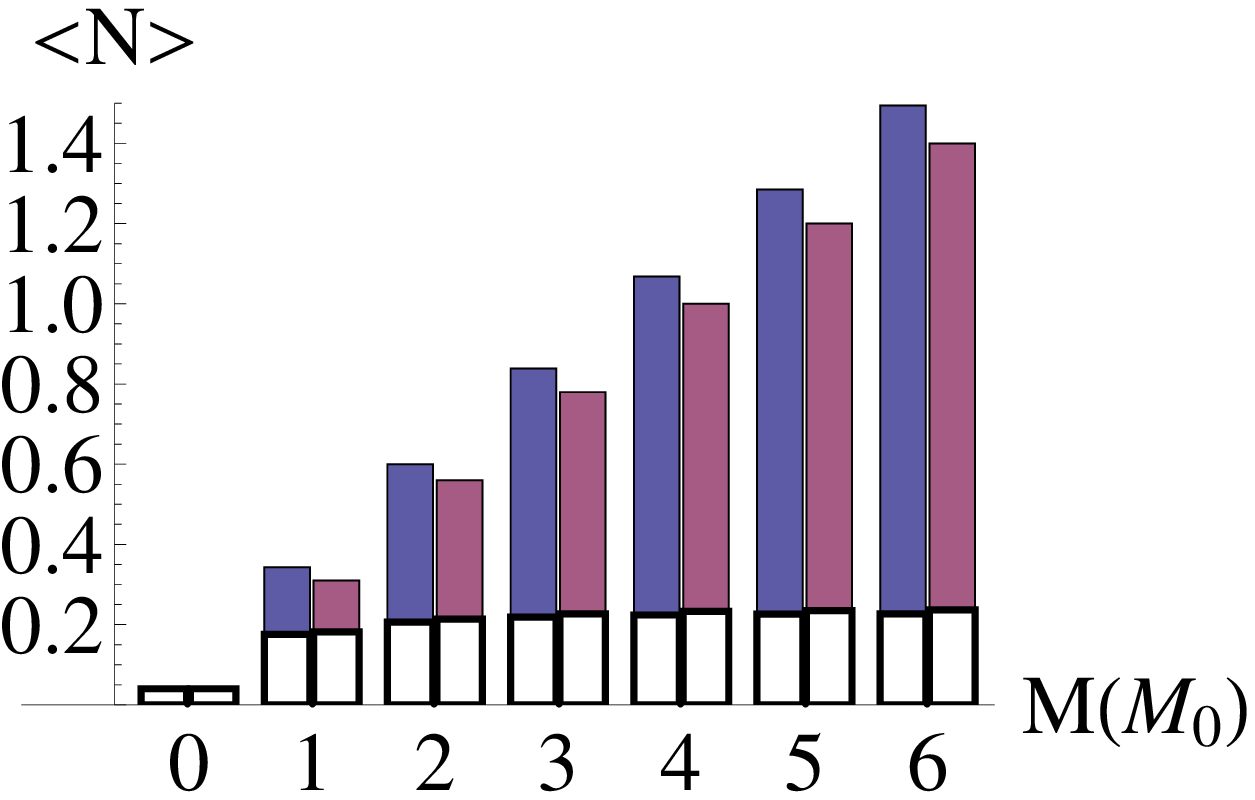,width=0.45\linewidth}}\vspace{0.5cm}
\textbf{(c)}\hskip3.5cm \textbf{(d)}\vskip-0.5cm
\centerline{\psfig{figure = 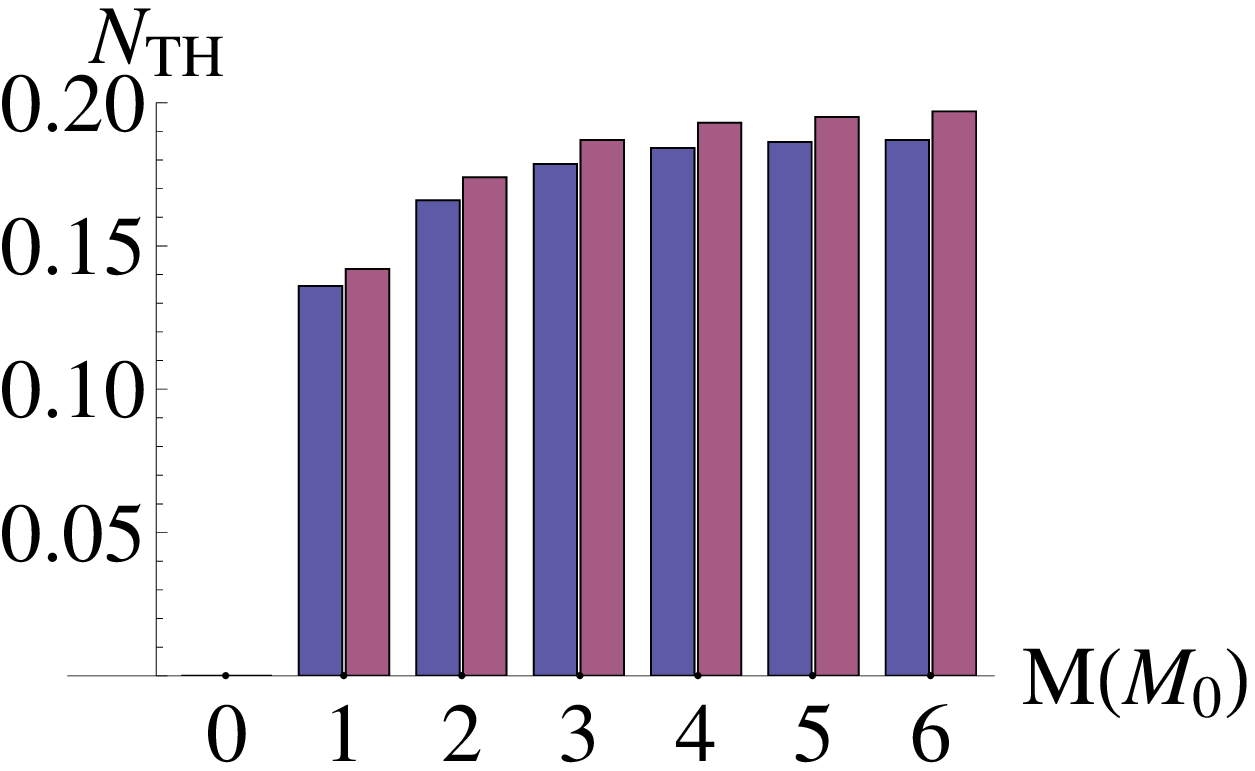,width=0.45\linewidth}\psfig{figure = 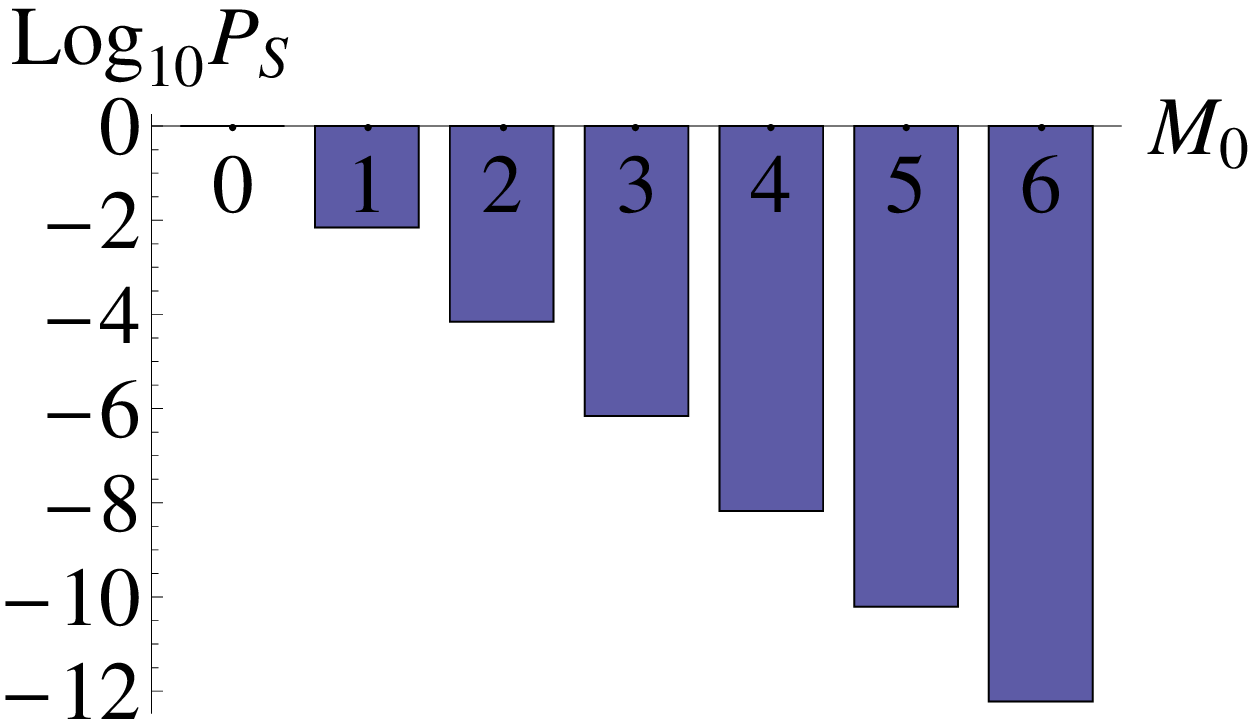,width=0.45\linewidth}}
\vskip0.3cm
\hskip-3.5cm\textbf{(e)}\vskip-0.5cm
\centerline{\psfig{figure = 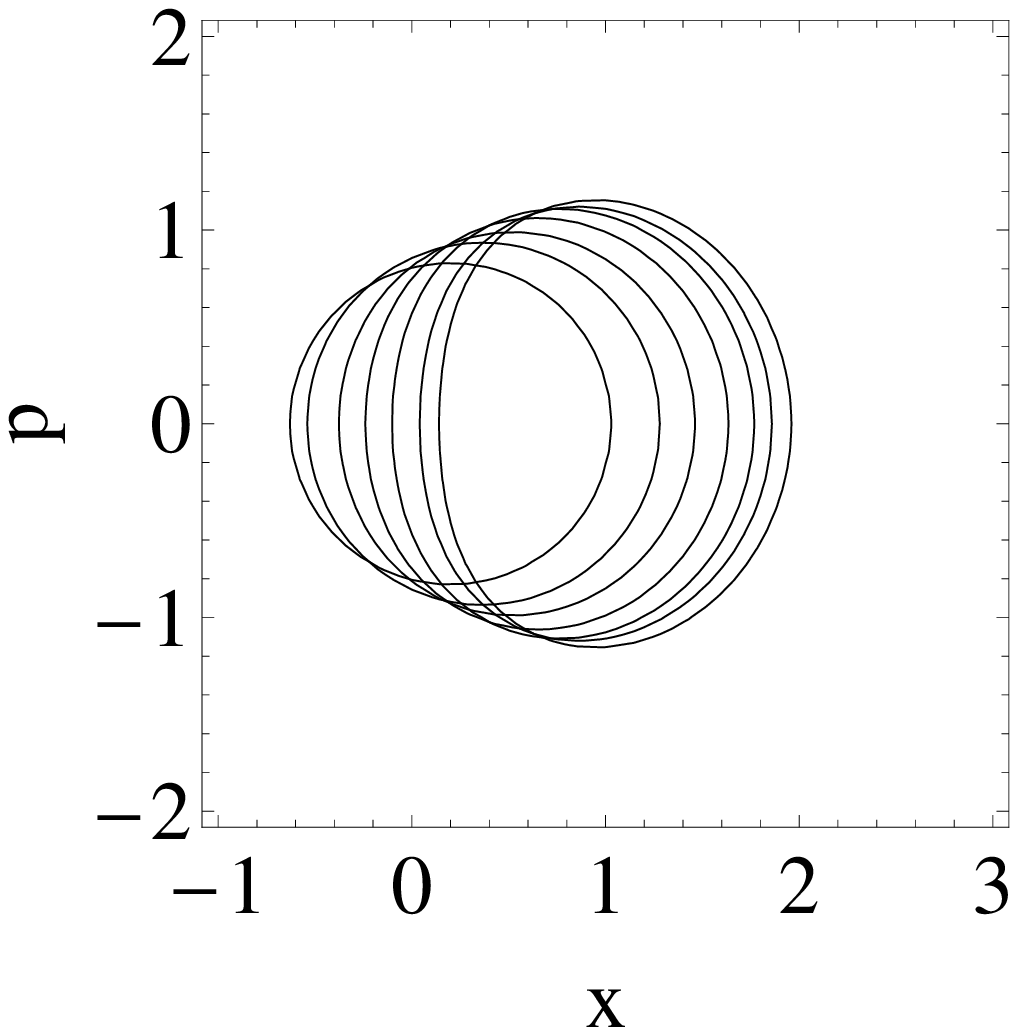,width=0.45\linewidth}}
\caption{Probabilistic phase concentration in highly nonlinear amplification. The canonical phase variance $V_C$ \textbf{(a)}, total mean photon number $\langle N\rangle$ \textbf{(b)}, optimal mean photon number of added thermal photons $N_{TH}$ \textbf{(c)}, the success rate of feasible photon subtraction \textbf{(d)} and contours plot of Wigner function for different number of subtractions $M=0,\ldots,6$ \textbf{(e)}, for the initial coherent state with $N=0.04$. The blue bars (left) correspond to the ideal subtraction of $M$ photons, the red bars (right) for feasible subtraction using beam splitter tap ($T=0.9$) and the threshold detector with efficiency $\eta=0.4$. The grey bars in \textbf{(a)} denote the variance for the ideal noiseless amplifier with gain $\langle N\rangle/N$.
The white bars in \textbf{(b)}  represent the total mean number of photons before the subtraction. The contours of Wigner functions in \textbf{(e)} go from left to right as $M=0,\ldots,6$.}
\label{fig_holevo}
\end{figure}
%%%%%%%%%%%%%%%%%%%%%%%%%%%%%%%%%%%%%%%%%%%%%%%%%%%%%%%%%%%%%%%%%%%%%%%%%%%%%%%%%%%%

For a coherent state transformed in this way, $aa^{\dag}|\alpha\rangle$, the total mean photon number $\langle N\rangle =N(4+5N+N^2)/(1+3N+N^2)$ increases and  the canonical variance obtained with help of
\begin{eqnarray}
    \mu_{C}=\exp(-N)\frac{\sqrt{N}}{1+3N+N^2}\sum_{n=0}^{\infty} \frac{N^n(n+1)(n+2)}{n!\sqrt{n+1}},
\end{eqnarray}
is always lower than the Holevo variance from (\ref{hol1}). For a lower $N<1$, the canonical phase variance after the probabilistic procedure approaches the phase variance for the coherent state with $N=\langle N\rangle$. For a larger $N$ this effect tends to be less pronounced as relative influence of single photon operations diminishes. In this scenario it is convenient to consider a generalization, collective $M$-photon addition followed by $M$-photon subtraction. The phase variance is then determined by
\begin{eqnarray}
    \mu_{C}&= &e^{-N}\frac{\sqrt{N}}{\cal{N}}\times\nonumber\\
    & &\sum_{n=0}^{\infty} \frac{N^n}{n!\sqrt{n+1}}\frac{(n+M)!}{n!}{(n+1+M)}{(n+1)!},\nonumber\\
    \cal{N}&=&e^{-N}\sum_{n=0}^{\infty} \frac{N^n}{n!}\left(\frac{(n+M)!}{n!}\right)^2,
\end{eqnarray}
and it decreases as $M$ grows. Simultaneously, this also leads to increase of the mean photon number.  For sufficiently low values of $N$ and $M$ the canonical variance approaches the result of the ideal noiseless amplifier and we can use the approximation
\begin{equation}\label{Maddsub}
V_{C}(N)\approx \frac{1}{(M+1)^2 N}+1-\frac{M+2}{\sqrt{2}(M+1)}+\mbox{O}^2(N).
\end{equation}
Comparison to the analogous formula for the noiseless amplifier (\ref{variances}) with $ N \rightarrow g^2 N$, reveals that
 $M+1$ can play a role of the amplification gain.

%%%%%%%%%%%%%%%%%%%%%%%%%%%%%%%%%%%%%%%%%%%%%%%%%%%%%%%%%%%%%%%%%%%%%%%%%%%%%
\begin{figure}
\textbf{(a)}\hskip3.5cm \textbf{(b)}
\vskip-0.5cm \centerline{\psfig{figure = 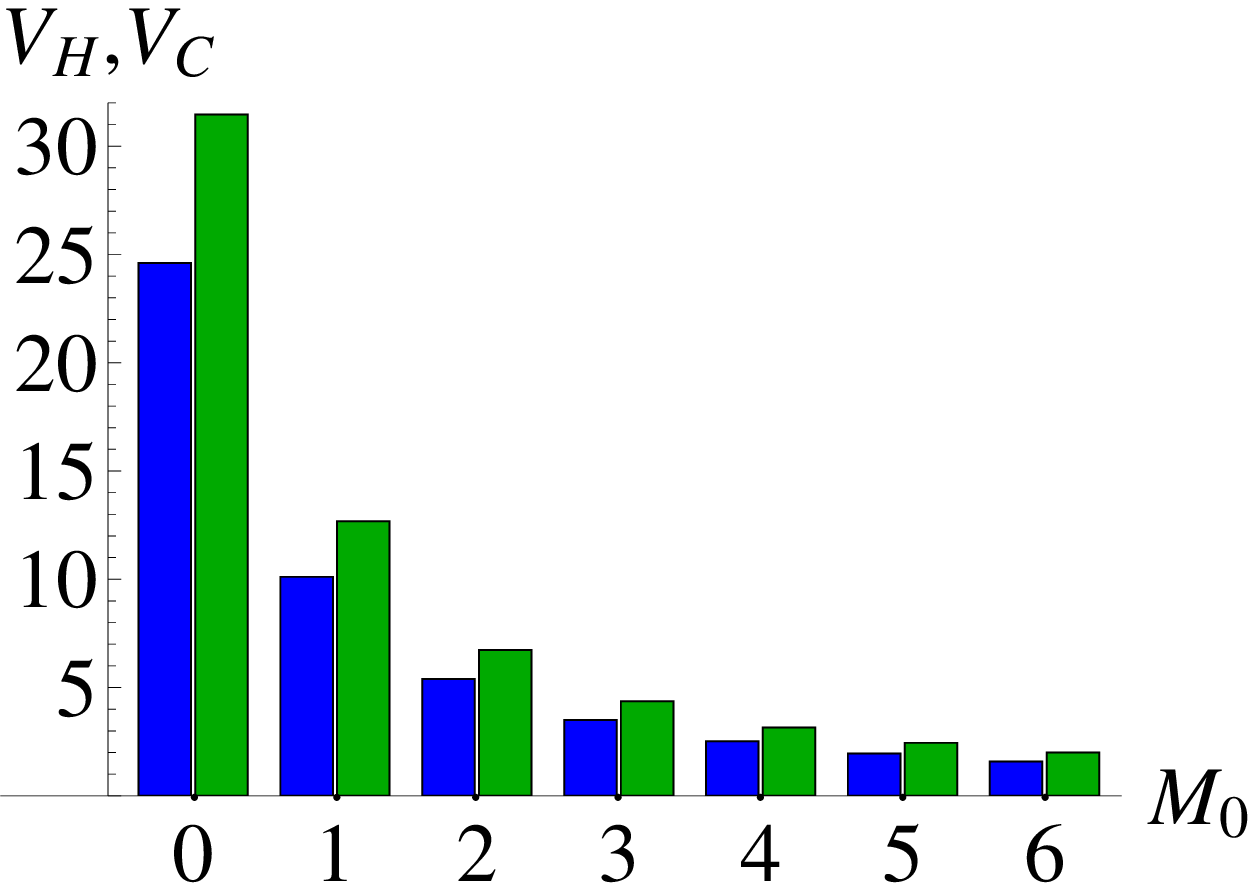,width=0.45\linewidth}\psfig{figure = 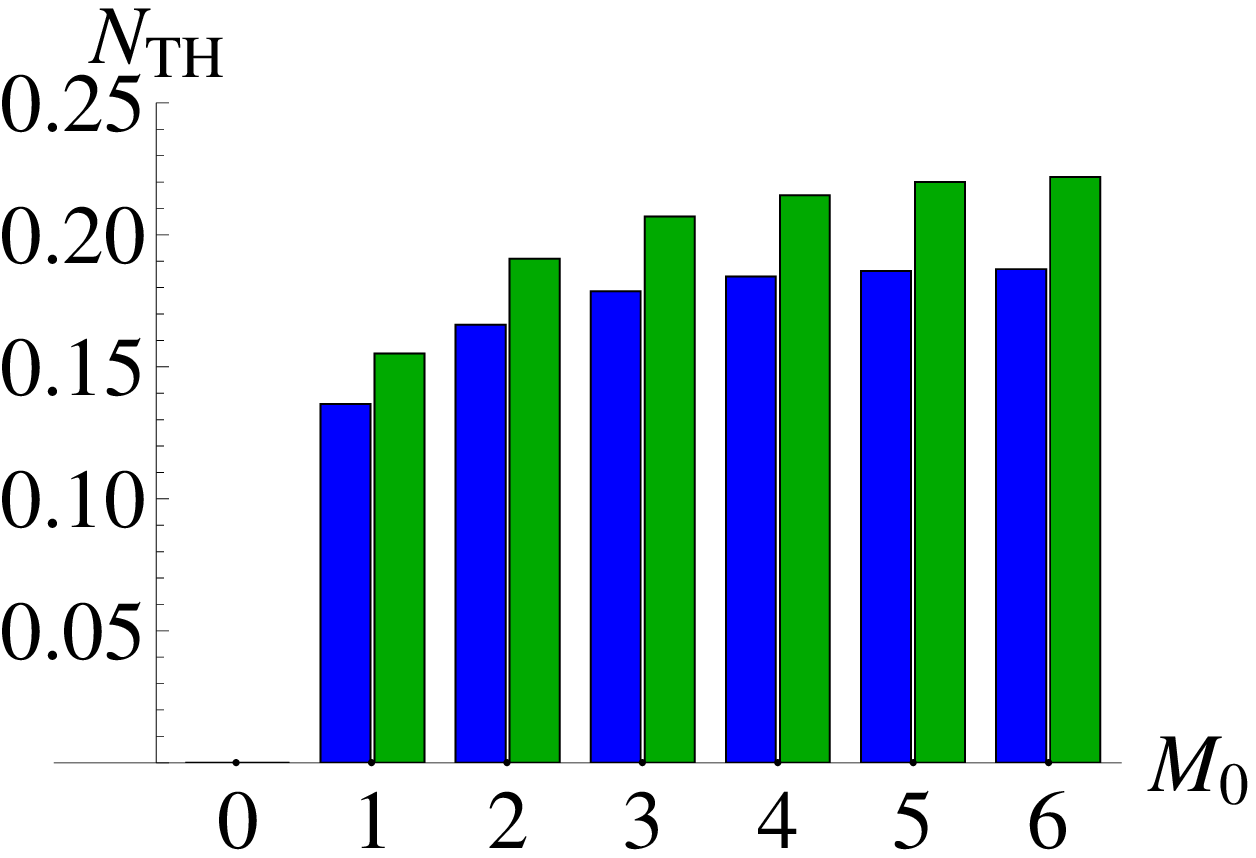,width=0.45\linewidth}}
\caption{Comparison between canonical phase measurement (blue bars, left) and variance of phase using the heterodyne measurement (green bars, right) after the realistic phase concentration discussed in Fig.~\ref{fig_holevo}.}
\label{fig_het}
\end{figure}
%%%%%%%%%%%%%%%%%%%%%%%%%%%%%%%%%%%%%%%%%%%%%%%%%%%%%%%%%%%%%%%%%%%%%%%%%%%%%%%%%%%%%%%%%%%%%%%%

For construction of such the probabilistic phase-insensitive amplifier a photon addition operation is required. Furthermore, the photons have to be added coherently, perfectly interfering with the incoming coherent state. This task can be performed using a non-degenerate optical parametric amplifier with avalanche photo-diode monitoring the output idler port \cite{photonaddition}. This approach has been already used to verify validity of commutation relations for annihilation operator \cite{commutation}, and it is therefore fully capable of demonstrating the probabilistic amplification for $M=1$. However, the procedure is not trivial and adding  and subsequently subtracting two or more photons is currently unfeasible, mainly due to low success rates.

Surprisingly, there is another effect we can take advantage of. Instead of adding single photons, we can add phase insensitive thermal noise characterized by its mean number of thermal photons $N_{TH}$. After the $M$-photon subtraction the density matrix of the state is
$
 \rho= \sum_{n,m} \rho_{n,m} |n-M\rangle\langle m-M|,
$
where
\begin{eqnarray}
\rho_{n,m}&=& \frac{1}{\cal{N}}\sqrt{\frac{n!}{m!}}\exp\left(-\frac{|\alpha|^2}{N_{TH}+1}\right)\frac{(\alpha^{*})^{m-n}N_{TH}^n}{(N_{TH}+1)^{m+1}}
\times\nonumber\\
& &\mbox{L}_{n}^{m-n}\left(
-\frac{|\alpha|^2}{N_{TH}(N_{TH}+1)}\right)\times\nonumber\\
& &\sqrt{\frac{n!m!}{(n-M)!(m-M)!}}
\end{eqnarray}
for $m\ge n$ and $\rho_{m,n} = \rho_{n,m}^*$ otherwise. The normalization factor representing the success rate is
\begin{eqnarray}
\cal{N}&=&\sum_{k}\exp\left(-\frac{|\alpha|^2}{N_{TH}+1}\right)\frac{N_{TH}^{k+M}}{(N_{TH}+1)^{k+M+1}}\times\nonumber\\
& &\mbox{L}_{k+M}^{0}\left(
-\frac{|\alpha|^2}{N_{TH}(N_{TH}+1)}\right)\frac{(k+M)!}{k!}.
\end{eqnarray}
The canonical phase variance can be calculated using
\begin{eqnarray}\label{noisesub}
\mu_{C}&=&\frac{1}{\cal{N}}\sum_{k}\sqrt{\frac{(k+M)!}{(k+M+1)!}}\exp\left(-\frac{|\alpha|^2}{N_{TH}+1}\right)
\times\nonumber\\
& &\frac{\alpha N_{TH}^{k+M}}{(N_{TH}+1)^{k+M+2}}\mbox{L}_{k+M}^{1}\left(
-\frac{|\alpha|^2}{N_{TH}(N_{TH}+1)}\right)\times\nonumber\\
& &\sqrt{\frac{(k+M)!(k+1+M)!}{k!(k+1)!}}.
\end{eqnarray}

As can be seen in Fig.~\ref{fig_holevo}, the probabilistic amplification of the initial coherent state ($M=0$) results in reduction of the phase variance and in increase of the total mean photon number $\langle N\rangle$ for both the ideal and realistic photon subtractions. The mean number $N_{TH}$ of added thermal photons was optimized to minimize the phase variance and it is saturating for larger $M$. The reduction of the phase variance saturates as well, but already for a feasible four photon subtraction the resulting phase variance corresponds to the phase variance of a coherent state with $N = 0.36$ (as opposed to the coherent state with $N=0.04$ before the amplification). This is equivalent to a strong amplification with gain $G=g^2=\langle N\rangle/N=9$.

The non-linear nature of the amplification is well visible from a change of the contour of Wigner function (taken at full width at a half maximum) in Fig.~\ref{fig_holevo}. The initial circular contour gains a "crescent" shape as $M$ and increases, which is a difference from the ideal noiseless amplification methods \cite{ampi1,ampi2}, which keep the state coherent. However, although the change of shape of the Wigner function suggests greater phase uncertainty, the increase of amplitude of the state results in smaller Holevo variance. This effect can be understood by realizing that the photon subtraction applied to a mixture of coherent states `picks' states with highest amplitudes, which in the case of the displaced thermal states are those in the radial direction.

For a physical understanding it is illustrative to consider a weak coherent state $|0\rangle+\alpha|1\rangle$ displaced by a weak thermal noise $\rho\rightarrow \rho+\epsilon_{TH}(a^{\dagger}\rho a+a\rho a^{\dagger})$ and followed by a single photon subtraction. The resulting state is $N|0\rangle\langle 0|+\epsilon_{TH}(|0\rangle+2\alpha|1\rangle)(\langle 0|+2\alpha^{*}\langle 1|)$ up to a normalization ${\cal N}=N+\epsilon_{TH}+4N\epsilon_{TH}$. The canonical phase variance can be determined from $\mu=2\epsilon_{TH}\alpha /\cal{N}$ and for small $N<0.1$, the reduction approaches $V\propto \frac{1}{4N}$, approximating very well the result for the ideal amplification (\ref{variances}) with $g=2$, if $\epsilon_{TH}$ is low enough.  More generally, approximating the thermal noise as adding up to $M$ photons followed by the $M$-photon subtraction leads to the phase variance  $V\propto \frac{1}{(M+1)^2N}$, which is qualitatively matching the results for ideal amplification with $g = M+1$ (\ref{variances}), as well as amplification by adding and subtracting $M$ photons (\ref{Maddsub}).

A feasible scheme capable to approximately subtract $M$ photons, which is required for physical implementation of the procedure, is sketched in Fig.~\ref{fig_setup}. It can be built using a linear coupling (beam splitter with the transmissivity $T$) to tap a part of optical signal and then implementing threshold measurement registering at least $M_{0}$ photons \cite{detect}. The quantum efficiency of the detector can be modeled by a virtual beam splitter with transmissivity $\eta$ inserted in front of thr ideal detector. The quality of the outgoing signal depends on the transmissivity $T$ - values $T<1$ translate as loss, which increases the phase variance. On the other hand, the limited quantum efficiency of the detector only affects the success rate. However, $\eta$ too low may require lower $T$ to achieve sufficiently high success rates.

Generally, the amplified state can be expressed as
\begin{eqnarray}\label{filtered}
    \rho_{out} &=& \frac{1}{P_S}\int
\Phi\left(\frac{\beta}{\sqrt{T}}\right)\mathcal{P}_{\Pi}\left(\frac{\beta}{\sqrt{T}}\right)
|\beta\rangle\langle\beta| \frac{d^2\beta}{T},
\end{eqnarray}
where $\mathcal{P}_{\Pi}(\beta) = \langle \sqrt{\eta(1-T)}\beta |\Pi|\sqrt{\eta(1-T)} \beta \rangle$ and $\Pi$ denotes the positive detection POVM element, which in case of the threshold detector looks as $\Pi = 1 - \sum_{k=0}^{M_0-1}|k\rangle\langle k |$.
The initial coherent state with addition of thermal noise is represented by $\Phi(\beta) = \exp({-|\beta-\alpha|^2/N_{\mathrm{th}}})/\pi N_{\mathrm{th}}$. The normalization factor $P_S$ gives probability of the success:
$P_S = \int \Phi(\beta/\sqrt{T})\mathcal{P}_{\Pi}(\beta/\sqrt{T})d^2\beta/T.$

Detailed analysis of the procedure is beyond the scope of this letter. However, it can be seen in Fig~\ref{fig_holevo} that the approximative multi-photon subtraction, even with low quantum efficiency, is a sufficient replacement for the ideal subtraction.
Furthermore, in Fig.~\ref{fig_het} the comparison of the canonical and the heterodyne variances shows a good qualitative agreement and justifies the use of the canonical measurement.

In summary, we have proposed the phase concentration of coherent states by a probabilistic measurement-induced amplifier. The amplifier setup based on thermal noise addition followed by feasible multi-photon subtraction allows to substantially reduce the phase variance of a coherent state. This allows to probabilistically correct loss in the channel, reducing need for higher amplitudes of transmitted coherent states. Further applications as probabilistic quantum cloning, entanglement manipulation, or coherent states quantum cryptography are also very attractive.

The research has been supported by
projects No. MSM 6198959213 and No. LC06007 of the
Czech Ministry of Education, grant 202/07/J040 of
GA CR and EU grant No. 212008, COMPAS. R.F. also acknowledges a support by the Alexander
von Humboldt Foundation.

\end{document}